\documentclass[12pt]{article}
\input epsf.tex
\usepackage{cite}
\usepackage{latexsym}
\usepackage{graphicx}
\usepackage{amsmath,amstext}
\usepackage{comment}
\numberwithin{equation}{section}
\newcommand{\be}{\begin{equation}}
\newcommand{\ee}{\end{equation}}
\newcommand{\benn}{\begin{equation*}}
\newcommand{\eenn}{\end{equation*}}
\newcommand{\bea}{\begin{eqnarray}}
\newcommand{\eea}{\end{eqnarray}}
\newcommand{\bean}{\begin{eqnarray*}}
\newcommand{\eean}{\end{eqnarray*}}

\def\centeron#1#2{{\setbox0=\hbox{#1}\setbox1=\hbox{#2}\ifdim
\wd1>\wd0\kern.5\wd1\kern-.5\wd0\fi
\copy0\kern-.5\wd0\kern-.5\wd1\copy1\ifdim\wd0>\wd1
\kern.5\wd0\kern-.5\wd1\fi}}
\def\ltap{\;\centeron{\raise.35ex\hbox{$<$}}{\lower.65ex\hbox{$\sim$}}\;}
\def\gtap{\;\centeron{\raise.35ex\hbox{$>$}}{\lower.65ex\hbox{$\sim$}}\;}

\begin{document}
\begin{titlepage}
\begin{center}
\hfill SCIPP-2008/RUNHETC \\

\vskip 0.2in

{\Large \bf Locality and the classical limit of quantum mechanics }

\vskip 0.3in

Tom Banks ,$^1$

\vskip 0.2in

\emph{$^1$ Santa Cruz Institute for Particle Physics,\\
     Santa Cruz CA 95064 and Rutgers NHETC, 126 Frelinghuysen Rd., \\ Piscataway,
     NJ, 08854}

\begin{abstract}
 I argue that conventional estimates of the criterion for
classical behavior of a macroscopic body are incorrect, because they
do not take into account the locality of interactions, which
characterizes the behavior of all systems described approximately by
local quantum field theory. Black holes are the only localized
objects which do not have such a description. The deviations from
classical behavior of a macroscopic body, except for those which can
be described as classical uncertainties in the initial values of
macroscopic variables,are {\it exponentially} small as a function of
the volume of the macro-system in microscopic units.
\end{abstract}

\end{center}
\end{titlepage}

\section{\bf Classical behavior in the non-relativistic quantum mechanics of particles}

In standard texts on non-relativistic quantum mechanics the
classical limit is described via examples and via the WKB
approximation. In particular, one often describes the spreading of
the wave packet of a free particle, and estimates it as a function
of time and the particle mass $M$. There is nothing wrong with the
mathematics done in these texts, but the implication that these
estimates provide the basis for an understanding of why classical
mechanics is such a good approximation for macroscopic objects is
not correct and therefore misleading. In particular it leads one to
conclude that the corrections to decoherence for a wave function
describing a superposition of two different macroscopic states is
power law in the mass. I would aver that this mistake forms part of
the psychological unease that many physicists feel about the
resolution of Schr\"{o}dinger's cat paradox in terms of the concept
of decoherence.

In this introductory section, I would like to demonstrate why such
estimates are wrong, using standard ideas of non-relativistic
quantum mechanics. In the remainder of the paper I will discuss the
basis for these calculations in quantum field theory. This will also
remove the necessity to resort to Hartree-Fock like approximations
to prove the point directly in the non-relativistic formalism. The
essential point of the argument is that we must take into account
the fact that a macroscopic object is made out of a huge number $
> 10^{20}$ of microscopic constituents, in order to truly understand
its classical behavior. I will argue that, as a consequence, the
overlaps between states where the object follows two macroscopically
different trajectories, as well as the matrix elements of all
local\footnote{In this context local means an operator which is a
sum of terms, each of which operates only on a few of the
constituent particles. A more precise, field theoretic, description
will be given in the next section.} operators between such states,
are of order
$$e^{ - 10^{20}} .$$ The extraordinary smallness of such double
exponentials defeats all of our ordinary intuitions about ordinary
physics. In a phrase I learned in graduate school from Kerson Huang,
``The number $e^{10^{20}}$, viewed as a time interval, is
essentially the same number when measured in Planck units, as it is
when measured in ages of the universe. The two differ by a mere
factor of $10^{61}$.". Over such long time scales, many
counter-intuitive things could happen. For example, in a
hypothetical classical model of a living organism made of this many
constituents, or in a correct quantum model, the phenomenon of
Poincare recurrences assures that given (roughly) this much time,the
organism could spontaneously self assemble,out of a generic initial
state of its constituents.

Consider then, the wave function of such a composite of $N \gg 1$
particles, assuming a Hamiltonian of the form
$$H = \sum \frac{\overrightarrow{p}_i^2}{ 2 m_i} + \sum V_{ij} (x_i -
x_j).$$ Apart from electromagnetic and gravitational forces, the two
body potentials are assumed to be short ranged. We could also add
multi-body potentials, as long as the number of particles that
interact is $\ll N$\footnote{Or that the strength of $k$ body
interactions fall off sufficiently rapidly with $k$ for $k > N_0 \ll
N$.}.

The Hamiltonian is Galilean invariant and we can separate it into
the kinetic energy of the center of mass, and the Hamiltonian for
the body at rest. The wave function is of the form

$$\psi (X_{cm}) \Psi (x_i - x_j) ,$$ which is a general function of coordinate
differences. We now want to compare this wave function with the
internal wave function of the system when the particle is not
following a straight, constant velocity trajectory. In order to do
this, we introduce an external potential $U(\{ x_i\} )$. It is {\it
extremely} important that $U$ is not simply a function of the center
of mass coordinate but a sum of terms denoting the interaction of
the potential with each of the constituents. This very natural
assumption is derivable from local field theory: the external
potential must interact locally with ``the field that creates a
particle at a point".  So we assume
$$U = \sum u_i(x_i),$$ where we have allowed for the possibility,
{\it e.g.} that the external field is electrical and different
constituents have different charge.

To solve the external potential problem, we write $x_i = X_{cm} +
\Delta_i$ and expand the individual potentials around the center of
mass, treating the remaining terms as a small perturbation. We then
obtain a Hamiltonian for the center of mass, which has a mass of
order $N$, as well as a potential of order $N$.  The large $N$ limit
is then the WKB limit for the center of mass motion. The residual
Hamiltonian for the internal wave function has small external
potential terms, whose coefficients depend on the center of mass
coordinate.

The Schrodinger equation for the center of mass motion thus has
solutions which are wave functions concentrated around a classical
trajectory $X_{cm} (t)$ of the center of mass, moving in the
potential $\sum u_i (X_{cm}) $. These wave functions will spread
with time in a way that depends on this potential. For example,
initial Gaussian wave packets for a free particle will have a width,
which behaves like $\sqrt{t/N m}$ for large $t$, where $m$ is a
microscopic mass scale. The fact that this is only significant when
$t \sim N$ is the conventional explanation for the classical
behavior of the center of mass variable.

In fact, this argument misses the crucial point, namely that the
small perturbation, which gives the Hamiltonian of the internal
structure a time dependence, through the appearance of $X_{cm} (t)$,
is not at all negligible. To illustrate this let us imagine that the
wave function at rest has the Hartree-Fock form, an anti-symmetrized
product of one body wave functions $\psi_i (x_i)$, and let us
characterize the external potential by a strength $\epsilon$. As a
consequence of the perturbation, each one body wave function will be
perturbed, and its overlap with the original one body wave function
will be less than one. {\it It follows that the overlap between the
perturbed and unperturbed multi-body wave functions will be of order
$(1 - \epsilon)^N$}.  This has the exponential suppression we
claimed, as long as $\epsilon \gg \frac{1}{N}$. It is easy to see
that a similar suppression obtains for matrix elements of few body
operators. One can argue that a similar suppression is obtained for
generalized Jastrow wave functions, with only few body correlations,
but a more general and convincing argument based on quantum field
theory will be given in the next section. Here we will follow
through the consequences of this exponential suppression.

The effect is to break up the full Hilbert space of the composite
object in the external potential, into {\it approximate
super-selection sectors} labeled by macroscopically different
classical trajectories $X_{cm} (t)$ (microscopically different
trajectories correspond to $\epsilon \sim {1\over N}$). Local
measurements cannot detect interference effects between states in
different super-selection sectors on times scales shorter than
$e^{10^{20}} $ (following Huang, we leave off the obviously
irrelevant unit of time).  That is to say, for all {\it in principle
purposes}, a superposition of states corresponding to different
classical trajectories behaves like a classical probability
distribution for classical trajectories. The difference of course is
that in classical statistical physics one avers that {\it in
principle} one could measure the initial conditions precisely,
whereas in quantum mechanics the uncertainty is intrinsic to the
formalism.

I have used the phrase {\it in principle} in two different ways in
the previous paragraph. The first use was ironic; the natural phrase
that comes to mind is {\it for all practical purposes}. I replace in
practice by in principle in order to emphasize that any conceivable
experiment that could distinguish between the classical probability
distribution and the quantum predictions would have to keep the
system isolated over times inconceivably longer than the age of the
universe. In other words, it is meaningless for a {\it physicist} to
consider the two calculations different from each other. In yet
another set of words; the phrase ``With enough effort, one can in
principle measure the quantum correlations in a superposition of
macroscopically different states", has the same status as the phrase
``If wishes were horses then beggars would ride".

The second use of {\it in principle} was the conventional
philosophical one: the mathematical formalism of classical
statistical mechanics contemplates arbitrarily precise measurements,
on which we superimpose a probability distribution which we
interpret to be a measure of our ignorance. In fact, even in
classical mechanics for a system whose entropy is order $10^{20}$,
this is arrant nonsense. The measurement of the precise state of
such a system would again take inconceivably longer than the age of
the universe.

This comparison is useful because it emphasizes the fact that the
tiny matrix elements between super-selection sectors are due to an
entropic effect. They are small because a change in the trajectory
of the center of mass changes the state of a huge number of degrees
of freedom. Indeed, in a very rough manner, one can say that the
time necessary to see quantum interference effects between two
macroscopically different states is of order the Poincare recurrence
time of the system. This is very rough, because there is no argument
that the order $1$ factors in the exponent are the same, so the
actual numbers could be vastly different. The important point is
that for truly macroscopic systems both times are
super-exponentially longer than the age of the universe.

The center of mass is one of a large number of {\it collective} or
{\it thermodynamic} observables of a typical macroscopic system
found in the laboratory. The number of such variables is a measure
of the number of macroscopic moving parts of the system. As we will
see, a system with a goodly supply of such moving parts is a good
measuring device. Indeed, the application of the foregoing remarks
to the quantum measurement problem is immediate. As von Neumann
first remarked, there is absolutely no problem in arranging a
unitary transformation which maps the state
$$\alpha |\uparrow \rangle > + \beta |\downarrow \rangle > \otimes |
Ready >,$$ of a microsystem uncorrelated with the $| Ready \rangle $
state of a measuring apparatus, into the correlated state
$$\alpha |\uparrow \rangle > \otimes | + \rangle + \beta |\downarrow \rangle > \otimes |
- >,$$ where $| +/- >$ are {\it pointer states} of the measuring
apparatus. If we simply assume, in accordance with experience, that
the labels $+/-$ characterize the value of a macroscopic observable
in the sense described above, then we can immediately come to the
following conclusions
\begin{itemize}

\item The quantum interference between the two pieces of the wave
function cannot be measured on time scales shorter than the
super-exponential times described above. The predictions of quantum
mechanics for this state are identical {\it in principle} (first
usage) to the predictions of a classical theory that tells us only
the probabilities of the machine reading $+$ or $-$. {\it Like any
such probabilistic theory} the algorithm for interpreting its
predictions is to condition the future predictions on any actual
measurements made at intermediate times. This is the famous
``collapse of the wave function", on which so much fatuous prose has
been expended. It no more violates conservation of probability than
does throwing out those weather simulations, which predicted that
Hurricane Katrina would hit Galveston.

\item One may worry that there is a violation of unitarity in this
description, because if I apply the {\it same} unitary
transformation to the states $|\uparrow \rangle \otimes | Ready
\rangle $ and $|\downarrow \rangle \otimes | Ready \rangle $,
individually, then I get a pair of states whose overlap is not
small. This seems like a violation of the superposition principle,
but this mathematical exercise has nothing to do with physics, for
at least two reasons. First the macro-states labeled by $+/-$ are
not single states, but huge ensembles, with $e^N$ members. The
typical member of any of these ensembles is a time dependent state
with the property that time averages of observables over a short
relaxation time are identical to those in another member of the
ensemble. The chances of starting with the identical $| Ready
\rangle$ state or ending with the same $ | +/- \rangle$ states in
two experiments with different initial micro-states, is $e^{-N}$.
Furthermore, and perhaps more importantly, the experimenter who
designs equipment to amplify microscopic signals into macroscopic
pointer readings, {\it does not} control the microscopic interaction
between the atoms in the measuring device and {\it e.g.} the
electron whose spin is being measured. Thus, in effect, every time
we do a new measurement, whether with the same input micro-state or
a different one, it is virtually certain that the unitary
transformation that is actually performed on the system is a
different one.

\end{itemize}

For me, these considerations resolve all the {\it angst} associated
with the Schr\"{o}dinger's cat paradox. Figurative superpositions of
live and dead cats occur every day, whenever a macroscopic event is
triggered by a micro-event. We see nothing remarkable about them
because quantum mechanics makes no remarkable predictions about
them. It never says ``the cat is both alive and dead", but rather,
``I can't predict whether the cat is alive or dead, only the
probability that you will find it alive or dead if you do the same
experiment over and over". Wave function collapse and the associated
claims of instantaneous action at a distance are really nothing but
the the familiar classical procedure of discarding those parts of a
probabilistic prediction, which are disproved by actual experiments.
This is usually called the use of conditional probabilities, and no
intellectual discomfort is attached to it.

We are left with the discomfort Einstein expressed in his famous
aphorism about mythical beings rolling dice. Those of us who
routinely think about the application of quantum mechanics to the
entire universe, as in the apparently successful inflationary
prediction of the nature of Cosmic Microwave Background temperature
fluctuations, cannot even find comfort in the frequentist's fairy
tale about defining probability ``objectively" by doing an infinite
number of experiments. Probability is a guess, a bet about the
future. What is it doing in the most precisely defined of sciences?
I will leave this question for each of my readers to ponder in
solitude. I certainly don't know the answer.

Finally, I want to return to the spread of the wave packet for the
center of mass, and what it means from the point of view presented
here. It is clear that the uncertainties described by this wave
function can all be attributed to the inevitable quantum
uncertainties in the initial conditions for the position and
velocity of this variable. Quantum mechanics prevents us from
isolating the initial phase space point with absolute precision. For
free particles or harmonic potentials, these can simply be viewed
(via the Wigner distribution) as microscopic initial uncertainties
in the classical trajectory $X_{cm} (t)$. For other potentials there
are small corrections to this, proportional to inverse powers of $N
\sim 10^{20}$, since the Wigner distribution no longer satisfies the
classical Liouville equation. If we wait long enough these
uncertainties would, from a purely classical point of view, lead to
macroscopic deviations of the position from that predicted by the
classical trajectory we have expanded around. The correct
interpretation of this is that our approximation breaks down over
such long time scales. A better approximation would be to decide
that after a time long enough for an initial microscopic deviation
to evolve into a macroscopic one, we must redefine our
super-selection sectors. After this time, matrix elements between
classical trajectories that were originally part of the same
super-selection sector, become so small that we must declare that
they are different sectors.

Thus instead of, in another famous Einsteinian phrase, complaining
that the moon is predicted to disappear when we don't look at it
(over a time scale power law in its mass), we say that quantum
mechanics predicts that our best measurement of the initial position
and velocity of the moon is imprecise. The initial uncertainties are
small, but grow with time, to the extent that eventually we cannot
predict exactly where the moon is. Quantum mechanics {\it does}
predict, that the moon has followed some (to a very good
approximation) classical trajectory, but does not allow us to say
which one, a long time after an initial measurement of the position
and velocity.

\section{Quantum field theory}

I will describe the considerations of this section in the language
of relativistic quantum field theory. {\it A fortiori} they apply to
the non-relativistic limit, which we discussed in first quantization
in the previous section. They also apply to cutoff field theories,
with some kind of spatial cutoff, like a space lattice. The key
property of all these systems is that the degrees of freedom are
labeled by points in a fixed spatial geometry, with a finite number
of canonical bosonic or fermionic variables per point. The
Hamiltonian of these degrees of freedom is a sum of terms, each of
which only couples together the points within a finite
radius\footnote{Various kinds of exponentially rapid falloff are
allowed, and would not effect the qualitative nature of our
results.} In the relativistic case of course the Hamiltonian is an
integral of a strictly local Hamiltonian density.

Let us first discuss the ground state of such a system. If the
theory has a mass gap, then the ground state expectation values of
products of local operators fall off exponentially beyond some
correlation length $L_c$. If $d$ is the spatial dimension of the
system,and $V$ is a volume $\gg L_c^d$, define the state
$$|\phi_c , V > ,$$ as the normalized state with minimum
expectation value of the Hamiltonian, subject to the constraint that
$$ \int_V d^d x\ \phi (x) / V = \Phi_c .$$  Let $v = V/L_c^d$. One can show, using the
assumption of a finite correlation length, that these states have
the following properties
\begin{itemize}

\item The quantum dynamics of the variable $\Phi_c$ is amenable to
the semi-classical approximation, with expansion parameter $\propto
1/v$.

\item The matrix elements of local operators between states with
different values of $\phi_c$ satisfy
$$ \langle \Phi_c , v | \phi_1 (x_1)\ldots \phi_n (x_n) |
\Phi_c^{\prime} , v \rangle \sim e^{-c v} ,
$$ where $n$ is kept finite as $v \rightarrow \infty$.

\item The interference terms in superpositions between states with different values of
$\Phi_c$ remain small for times of order $e^{b v}$. This follows
from the previous remark and the fact that the Hamiltonian is an
integral of local operators. This remark is proved by thinking about
which term in the t-expansion of $e^{-iHt}$ first links together the
different superposition sectors with an amplitude of order $1$.
There is a technical problem in this argument, because the
Hamiltonian is unbounded, but it is intuitively clear that a cutoff
at high energy should not affect the infrared considerations here.

\end{itemize}

In the language of the previous section, {\it averages of local
fields over distances large compared to the correlation length} are
good pointer observables.

To define an actual apparatus, we have to assume that the quantum
field theory admits bound states of arbitrarily large size.
Typically this might require us to add chemical potential terms to
the Hamiltonian and insist on macroscopically large expectation
values for some conserved charge.  The canonical example would be a
large, but finite, volume drop of nuclear matter in QCD.  We can
repeat the discussion above for averages over sub-volumes of the
droplet.

Of course, in the real world, the assumption of a microscopically
small correlation length is not valid, because of electromagnetic
and gravitational forces. Indeed, most real measuring devices use
these long range forces, both to stabilize the bound state and for
the operation of the machine itself. I do not know how to provide a
mathematical proof, but I am confident that the properties described
above survive without qualitative modification\footnote{In the
intuitive physics sense, not that of mathematical rigor.}. This is
probably because all the long range quantum correlations are
summarized by the classical electromagnetic and gravitational
interactions between parts of the system\footnote{ Recall that the
Coulomb and Newtonian forces between localized sources are described
in quantum field theory as quantum phase correlations in the wave
function for the multi-source system.} . It would be desirable to
have a better understanding of the modification of the arguments
given here, which is necessary to incorporate the effects of
electromagnetism and (perturbative) gravitation. One may also
conclude from this discussion that a system at a quantum critical
point, which has long range correlations not attributable to
electromagnetism or gravitation, would make a poor measuring device,
and might be the best candidate for seeing quantum interference
between ``macroscopic objects".  Of course, such conformally
invariant systems do not have large bound states which could serve
as candidate ``macroscopic objects".

Despite the mention of gravitation in the previous paragraph, the
above remarks do not apply to regimes in which the correct theory of
quantum gravity is necessary for a correct description of nature. We
are far from a complete understanding of a quantum theory of
gravity, but this author believes that it is definitely not a
quantum field theory. I will give a necessarily idiosyncratic
discussion of quantum gravity in the next section. The reader who
wishes to skip that section will not lose anything essential to his
understanding of the general principles of quantum theory.
\section{Towards a quantum theory of gravitation}

This is not the place to recapitulate all I have written about the
quantum theory of gravitation. Readers who are interested can
consult \cite{qg}.  Instead, I will just assert certain differences
between my vision of the theory, and ordinary quantum field theory.
In quantum field theory, one assumes a fixed Lorentzian space-time
background, and the algebra of observables describing the results of
measurements in a given causal diamond\footnote{A causal diamond is
the intersection of interior of the past light cone of a point P and
the interior of the future light cone of a point Q in the past of
P.}, is infinite dimensional and becomes universal as the diamond is
taken smaller and smaller. This is a fancy way of saying that all
quantum field theories are scale invariant at short distance.

The covariant entropy bound \cite{fsb} is a conjectural bound on the
entropy associated with a causal diamond in the theory of quantum
gravity. It says that the entropy is bounded by one quarter of the
area in Planck units, of the maximal area $d-2$ surface on the null
boundary of the diamond. This maximal area surface is called the
{\it holographic screen} of the causal diamond. The covariant
entropy bound can be ``derived" in a semi-classical manner by {\it
assuming} a connection between entropy density and energy density.
Fischler and I\cite{holocosm} proposed that this connection between
geometry and entropy be taken as the fundamental constructive
principle of quantum gravity. The only universally defined density
matrix for a quantum system, which, like quantum field theory in a
generic space-time, does not necessarily have a conserved
Hamiltonian,is the maximally uncertain one. We thus proposed that
the quantum definition of a causal diamond is a Hilbert space of
fixed size. The causal relations between different diamonds are
fixed by specifying which tensor factors in the algebra of
observables in each diamond, should be identified as operators which
can be measured by observers in both diamonds. We presented a
tentative set of axioms for such a quantum space-time\cite{qg}.

For the present purposes, the most important feature of this
formalism is the (only partially understood) manner in which it
reduces to quantum field theory, and its description of high entropy
states which are not approximated by quantum field theory. Very
roughly speaking, the variables describing physics in a causal
diamond can be described by first specifying the ``algebra of
functions on the holographic screen". This is a finite dimensional
matrix algebra, with a basis consisting of $N$ elements. When $N
\rightarrow\infty$ the area of the screen goes to infinity and the
algebra converges to the usual infinite dimensional commutative
algebra of functions. A basis element $f_n$ of the algebra is called
a pixel. The variables of the quantum theory form an operator
algebra
$$\otimes_{\cal P} {\cal A} ,$$ where the single pixel algebra ${\cal
A }$ is a finite dimensional operator algebra\footnote{This algebra
incorporates information about the detailed topology of space-time,
and in particular, compact dimensions.}. In less formal language, we
pixelate the holographic screen and have an independent algebra of
quantum variables for each pixel.

The quantum variables then are matrices $O_i^j$, where for each
element $i,j$, $O$ is a member of ${\cal A}$. The precise algebra of
matrices is determined by the requirement that the sequence of
algebras converge to the correct continuum function algebra as $N
\rightarrow\infty$. Particles arise in this formalism when, for
dynamical reasons, the variables associated with a commuting set of
matrices, are approximately dynamically independent of the rest. The
usual permutation gauge symmetry relating commuting matrices to
block diagonal matrices is interpreted as particle statistics. There
is a natural way\cite{susholo} that the geometry of a holographic
screen produces variables describing {\it supersymmetric} particles.
As in Matrix Theory\cite{bfss}, the size of individual block
matrices is related to the total momentum of particles in the
direction perpendicular to the holographic screen. We will call this
radial momentum. The natural basis for particle kinematics is to
describe each particle by an angular position on the screen and the
momentum perpendicular to the screen. This should be familiar to
experimental particle physicists.

Without going into details, two features of this formalism are
noteworthy:

\begin{itemize}

\item In a finite causal diamond, the number of particles is
bounded, in a way that depends on their momenta.

\item If we try to use all of the variables in a finite causal
diamond, we inevitably have non-commuting matrices, and we lose the
particle interpretation of the Hilbert space. This failure occurs as
we try to construct states which maximize the entropy\footnote{As
usual, this locution means, "states which are typical members of the
maximal entropy ensemble".}.  A further assumption of the formalism
is that the Hamiltonian for these maximal entropy states is a highly
degenerate one with a random spectrum. This ensures that our model
reproduces the thermodynamics of black holes.

\end{itemize}

To make a long and incomplete story short, the aim of this formalism
is to show that the theory of quantum gravity contains states
consisting of multiple particles, under conditions in which the
particles do not collapse to form black holes, as well as black hole
states, {\it and nothing else}\footnote{I am neglecting the states
on the cosmological horizon of an asymptotically de Sitter
space-time, which behave much like black hole states.}. The black
hole states have tiny random energy splittings, and behave like an
equilibrium ensemble with only a few macroscopic variables. The
particle states are well described by field theory and, for
appropriate choices of field theory\footnote{Presumably these
choices are somewhat restricted in the quantum theory of gravity,
and existing approaches to string theory have shown us some of these
restrictions. However, as of this writing, a wide range appears to
be possible.} it will have macroscopic bound states with many
macroscopic moving parts. These are excellent approximations to the
classical measuring devices used in formulating the axioms of
quantum mechanics, as has been shown above. Black holes are terrible
measuring devices, they can only register changes in a few
macroscopic variables, like mass charge and angular momentum.

\section{Conclusions}

I suspect the material in this paper is well understood by many
other physicists, including most of those who have worked on the
environmental decoherence approach to quantum
measurement\cite{decoh}. If there is anything at all new in what I
have written here about quantum measurement, it lies in the
statement that a macroscopic apparatus of modest size serves as its
own ``environment" for the purpose of environmental decoherence. In
normal laboratory circumstances, the apparatus interacts with a much
larger environment and the huge recurrence and coherence times
become even larger. Nonetheless, there is no reason to suppose that
a modestly macroscopic apparatus, surrounded by a huge region of
vacuum, with the latter protected from external penetrating
radiation by thousands of meters of lead, would behave differently
over actual experimental time scales, than an identical piece of
machinery in the laboratory.

The essential point in this paper is that the corrections to the
classical behavior of macroscopic systems are exponential in the
size of the system in microscopic units. This puts observable
quantum behavior of these systems in the realm of Poincare
recurrence phenomenon, essentially a realm of science fiction rather
than of real experimental science. When a prediction of a scientific
theory can only be verified by experiments done over times
super-exponentially longer than the measured age of the universe,
one should not be surprised if that prediction is counter-intuitive
or ``defies ordinary logic".

Quantum mechanics does make predictions for macro-systems which are
different than those of deterministic classical physics. Any time a
macro-system is put into correlation with a microscopic variable -
and this is the essence of the measurement process - its behavior
becomes unpredictable. However, these predictions are
indistinguishable from those of classical statistical mechanics,
with a probability distribution for initial conditions derived from
the quantum mechanics of the micro-system. It is only if we try to
interpret this in terms of a classical model of the micro-system
that we realize something truly strange is going on. The predictions
of quantum mechanics {\it for micro-systems} {\it are} strange, and
defy the ordinary rules of logic. But they do obey a perfectly
consistent set of axioms of their own, and we have no real right to
expect the world beyond the direct ken of our senses, which had no
direct effect on the evolution of our brains, to "make sense" in
terms of the rules which were evolved to help us survive in a world
of macroscopic objects.

Many physicists, with full understanding of all these issues, will
still share Einstein's unease with an intrinsically probabilistic
theory of nature. Probability is, especially when applied to
non-reproducible phenomena like the universe as a whole, a theory of
guessing, and implicitly posits a mind which is doing the guessing.
Yet all of modern science seems to point in the direction of mind
and consciousness being an emergent phenomenon; a property of large
complex systems rather than of the fundamental microscopic laws. The
frequentist approach to probability does not really solve this
problem. Its precise predictions are only for fictional infinite
ensembles of experiments. If, after the millionth toss of a
supposedly fair coin has shown us a million heads, and we ask the
frequentist if we're being cheated, all he can answer is
``probably". Neither can he give us any better than even odds that
the next coin will come up tails if the coin toss is truly unbiased.

I have no real answer to this unease, other than ``That's life. Get
over it."  For me the beautiful way in which linear algebra
generates a new kind of probability theory, even if we choose to
ignore it and declare it illogical\footnote{One can easily imagine
an alternate universe, in which a gifted mathematician discovered
the non-commutative probability theory of quantum mechanics, and
speculated that it might have some application to real measurements,
long before experimental science discovered quantum mechanics.}, is
some solace for being faced with a question to which, perhaps, my
intrinsic makeup prevents me from getting an intuitively satisfying
answer. On the other hand, I believe that discomfort with an
intrinsically probabilistic formulation of fundamental laws is the
only ``mystery" of quantum mechanics.  If someone told me that the
fundamental theory of the world was classical mechanics, with a
fixed initial probability distribution, I would feel equally
uncomfortable. The fact that the laws of probability for
micro-systems don't obey our macroscopic ``logic" points only to
facts about the forces driving the evolution of our brains. If we
had needed an intuitive understanding of quantum mechanics to obtain
an adaptive advantage over frogs, we, or some other organism, would
have developed it. Perhaps we can breed humans who have such an
intuitive understanding by making the right to reproduce contingent
upon obtaining tenure at a physics department. Verifying the truth
of this conjecture would take a long time, but much less than time
than it would take to observe quantum correlations in a
superposition of macro-states.

\section{Acknowledgments}

I would like to thank Michael Nauenberg, Anthony Aguirre, Michael
Dine and Bruce Rosenblum, for important conversations.

This research was supported in part by DOE grant number
DE-FG03-92ER40689.

\end{document}